\documentclass[12pt,onecolumn,oneside,a4paper,peerreview]{IEEEtran}
\usepackage{cite}
\usepackage{graphicx}
\usepackage{amsmath}
\usepackage{times}
\usepackage{latexsym}
\usepackage{graphicx}
\usepackage{bm}
\usepackage{amssymb}
\usepackage{stfloats}
\usepackage{cases}
\usepackage{array}
\usepackage{setspace}
\usepackage{fancyhdr}
\usepackage{citesort}
\usepackage{multirow}

\usepackage{psfrag}
\usepackage{subfigure}
\usepackage{url}
\usepackage{epsfig}
\usepackage{caption2}
\usepackage{wrapfig}

\renewcommand{\QED}{\QEDopen}
\newtheorem{theorem}{Theorem}
\newtheorem{lemma}{Lemma}
\newtheorem{corollary}{Corollary}

\def\proof{\noindent\hspace{2em}{\itshape Proof: }}
\def\endproof{\hspace*{\fill}~$\square$\par\endtrivlist\unskip}

\begin{document}

\title{Outage Probability of Dual-Hop Multiple Antenna AF Relaying Systems with Interference}
\author{
\begin{minipage}{0.98\columnwidth}
\vspace*{1cm}
\begin{center}
\authorblockN{Caijun Zhong\authorrefmark{4},
              Himal A. Suraweera\authorrefmark{3},
              Aiping Huang\authorrefmark{4},
              Zhaoyang Zhang\authorrefmark{4}, Chau Yuen\authorrefmark{3},\\
              }
{\small \authorblockA{\authorrefmark{4}Institute of Information and Communication Engineering, Zhejiang University, Hangzhou, China}\\
\authorblockA{\authorrefmark{3}Engineering Product Development, Singapore University of Technology and Design, Singapore} \\
\vspace{-0.4cm}
\authorblockA{Email: \{caijunzhong, aiping.huang, ning\_ming\}@zju.edu.cn, \{himalsuraweera, yuenchau\}@sutd.edu.sg}}
\end{center}
\end{minipage}
}
\maketitle

\begin{abstract}
This paper presents an analytical investigation on the outage
performance of dual-hop multiple antenna amplify-and-forward
relaying systems in the presence of interference. For both the
fixed-gain and variable-gain relaying schemes, exact analytical
expressions for the outage probability of the systems are derived.
Moreover, simple outage probability approximations at the high
signal to noise ratio regime are provided, and the diversity order
achieved by the systems are characterized. Our results suggest that
variable-gain relaying systems always outperform the corresponding
fixed-gain relaying systems. In addition, the fixed-gain relaying
schemes only achieve diversity order of one, while the achievable
diversity order of the variable-gain relaying scheme depends on the
location of the multiple antennas.
\end{abstract}

\section{Introduction}
Due to the ability of significantly improving the throughput, coverage, and energy consumption of the communications systems, dual-hop
relaying technique has attracted enormous attention from both the industry \cite{3gpp} and
academia\cite{M.Hasna1,M.Hasna}. Among various relaying schemes proposed in the literature, amplify-and-forward (AF)
relaying scheme, which simply amplifies the received signal and re-transmits it to the destination, is of particular
interest because of its simplicity and low implementation cost.

The AF relaying scheme generally falls into two categories, i.e., fixed-gain relaying \cite{M.Hasna1} and variable-gain
relaying \cite{M.Hasna}. Both schemes have received great attention and a large body of literatures has investigated the
performance of the two relaying schemes in various propagation environments (see \cite{T.Tsiftsis,H.Suraw} and references therein). While these works have
significantly improved our understanding on the performance of dual-hop AF relaying systems, a key feature of wireless
communication systems, namely, co-channel interference (CCI), is neglected.

This important observation has recently promoted a surge of research
interest in understanding the impact of CCI on the performance of
dual-hop AF relaying systems. In \cite{C.Zhong}, the outage
performance of dual-hop fixed-gain AF relaying systems with
interference-limited destination was investigated, while
\cite{H.Suraweera} addressed case with variable-gain relaying scheme
and interference-limited relay, and later \cite{Fawaz} extended the
analysis of \cite{H.Suraweera} to the more general Nakagami-$m$
fading channels, while \cite{H.Sura2} studied the performance of
fixed-gain dual-hop systems with a Rician interferer. Meanwhile, the
more general case with interference at both the relay and
destination nodes has been investigated in
\cite{W.Xu,S.Chen,D.Lee1,S.Ikki0,AM,D.Costa}. In \cite{W.Xu}, the
outage performance of dual-hop fixed-gain AF relaying scheme was
examined, and the case with variable-gain relaying scheme was dealt
with in \cite{S.Chen,D.Lee1}. \cite{S.Ikki0} presented an
approximated error analysis of the system employing variable-gain
relaying scheme, and \cite{AM} studied the outage performance of
both fixed-gain and variable-gain schemes assuming a single dominant
interferer at both the relay and destination, while \cite{D.Costa,Himal12}
addressed the case with Nakagami-m fading. Most recently, resource
allocation problems in the AF relaying systems have been studied in
\cite{A.Nasri,S.Ikki2}.

It is worth pointing out that most of the prior works assume the
interference-limited scenario, hence, the impact of the joint effect
of CCI and noise on the outage performance of dual-hop AF relaying
system has not been well-understood. In addition, all the prior
works consider the single antenna systems, therefore, the effect of
employing multiple antennas in the presence of CCI in the dual-hop
context remains unknown. In light of these two key observations, we
investigate the outage performance of dual-hop multiple antenna AF
relaying systems in the presence of CCI as well as noise. For
mathematical tractability, we limit the analysis to the case where
only one of the nodes is equipped with multiple antennas, hence,
three scenarios are of interest: (1) multiple antenna source, single
antenna relay and destination ($\mbox{N}$-1-1); (2) multiple antenna
destination, single antenna source and relay (1-1-$\mbox{N}$); (3)
multiple antenna relay, single antenna source and destination
(1-$\mbox{N}$-1). We also assume that the relay node is subjected to
a single dominant interferer and noise while the destination is
corrupted by the noise only\footnote{Our model assumes that the
relay and destination nodes experience different patterns of
interference, which is particularly suitable for the frequency
division duplex system, where the source-relay link and the
relay-destination link operate over different
frequency\cite{H.Zheng,J.Gora}.}. Although the system model is less
general, it enables us to gain key design insights on the joint
impact of CCI and noise, as well as the benefit of implementing
multiple antennas.

The main contributions of the paper are summarised as follows:
\begin{itemize}
\item For fixed-gain relaying systems, we derive exact closed-form expressions for the outage probability of all three systems.
\item For variable-gain relaying systems, we present analytical expressions involving a single integral for the outage probability of all three systems. In addition, we propose simple and tight closed-form lower bound of the outage
probability of the system.
\item For both fixed-gain and variable-gain relaying systems, we give simple and informative high signal to noise ratio (SNR) approximations for the outage probability for all three systems.
\item These analytical expressions not only provide fast and efficient means for the evaluation of the outage performance of the systems, they also enable us to gain valuable insights on the impact of key system parameters on the outage performance of the
system.
\end{itemize}

The remaining of the paper is organized as follows: Section II introduces the system model. Section III presents the
exact as well as asymptotical analytical expressions for the outage probability of the systems, and numerical results
and discussions are provided in Section IV. Finally, Section V concludes the paper and summarizes the findings.

Notations: We use bold upper case letters to denote matrices, bold lower case letters to denote vectors and lower case
letters to denote scalers. $|{\bf h}|^2$ denotes the Frobenius norm, ${\tt E}\{x\}$ stands for the expectation of
random variable $x$, ${*}$ denotes the conjugate operator, while ${\dag}$ denotes the conjugate transpose operator.
$n!$ denotes the factorial of integer $n$ and $\Gamma(x)$ is the gamma function.

\section{System Model}
Let us consider a dual-hop multiple antenna AF relaying system as
illustrated in Figure 1, where the relay node is subjected to a
single dominate interferer\footnote{We assume a single interferer at
the relay node for the mathematical tractability. Although less
general, the single interferer model is still of practical interest
and importance. For instance, in a well planned cellular network, it
is very likely that the system will be subjected to a single
dominant interferer. Hence, it has been adopted in a number of
previous works, i.e., \cite{A.Shah,T.Poon}.} and additive white
Gaussian noise (AWGN), while the destination node is corrupted by
AWGN only.

During the first phase, the source transmits signal symbol to the relay node, and the signal received at the relay node
can be expressed as
\begin{align}
\mathbf{y}_r = \mathbf{H}_1{\bf x} +\mathbf{h}_Is_I+\mathbf{n}_1,
\end{align}
where ${\bf H}_1$ denotes the channel for the source-relay link, and its entries follow identically and independently
distributed (i.i.d.) complex Gaussian distribution with zero-mean and variance $\sigma_1^2$, ${\bf x}$ is the source
symbol vector with ${\tt E}\{{\bf x}^{\dag}{\bf x}\} = P$, ${\bf h}_I$ is the channel for the interference-relay link,
and its entries are i.i.d. complex Gaussian random variables with zero-mean and variance $\sigma_I^2$, $s_I$ is the
interference symbol satisfying ${\tt E}\{s_Is_I^{*}\} = P_I$, and $\mathbf{n}_1$ is the AWGN noise at the relay node
with ${\tt E}\{\mathbf{n}_1^{\dag}\mathbf{n}_1\} = N_0{\mathbf{I}}$.

At the second phase, the relay node transmits a transformed version of the received signal to the destination, and the signal at the destination can be expressed as
\begin{align}
{\bf y}_d = \mathbf{H}_2^{\dag} {\mathbf{W}}\mathbf{y}_r +{\bf n}_2,
\end{align}
where ${\bf H}_2$ denotes the channel for the relay-destination link, and its entries are i.i.d. complex Gaussian
random variables with zero-mean and variance $\sigma_2^2$. $\mathbf{W}$ is the transformation matrix with ${\tt
E}\{||\mathbf{W}\mathbf{y}_r||_F^2\} = P_r$, and ${\bf n}_2$ is the AWGN with ${\tt E}\{{\bf n}_2{\bf n}_2^{\dag}\} =
N_0{\bf I}$. We assume that ${\bf H}_1$, ${\bf H}_2$, and ${\bf h}_I$ are mutually independent.

Note, for the sake of a concise presentation, we have, in the above, provided a fairly general dual-hop multiple
antenna AF relaying system model on purpose. However, we do not specify the size of the matrices $\mathbf{H}_1$,
$\mathbf{H}_2$ or vectors $\mathbf{h}_I$ here. Instead, they will be defined explicitly whenever appropriate in the
following sections.

For notational convenience, we define $\rho_1 \triangleq \frac{P\sigma_1^2}{N_0}$, $\rho_2\triangleq
\frac{P_r\sigma_2^2}{N_0}$, and $\rho_I\triangleq \frac{P_I\sigma_I^2}{N_0}$.

\section{The $\mbox{N}$-$1$-$1$ System}
This section considers the case where the source node is equipped with $N$ antennas, while the relay and destination
nodes only have a single antenna. For such system, we assume that beamforming scheme is adopted at the source node,
i.e., ${\bf x} = {\bf w}_ts$, where ${\bf w}_t$ is the transmit beamforming vector with $|{\bf w}_t|^2 =1$ and $s$ is
the transmit symbol with ${\tt E}\{ss^*\} =P$. To conform to the notation convention, we will use vector ${\bf h}_1\sim
{\cal CN}^{1\times N}$ to denote the source-relay link instead ${\bf H}_1$. Similarly $h_I$ and $h_2$, will be used to
denote channel for the interference-relay link and relay-destination link, respectively.

The end-to-end signal to interference and noise ratio (SINR) of the system can be expressed as:
\begin{align}
\gamma =\frac{|w|^2|h_2|^2|{\bf h}_1{\bf w}_t|^2P}{|w|^2|h_2|^2|h_I|^2P_I+|w|^2|h_2|^2N_0+N_0}.
\end{align}
It is easy to observe that the optimal beamforming vector is to match the first hop channel ${\bf h}_1$, i.e., ${\bf
w}_t = \frac{{\bf h}_1^{\dag}}{|{\bf h}_1|}$. Therefore, the end-to-end SINR is given by
\begin{align}\label{eqn:sinr1}
\gamma =\frac{|w|^2|h_2|^2|{\bf h}_1|^2P}{|w|^2|h_2|^2|h_I|^2P_I+|w|^2|h_2|^2N_0+N_0}.
\end{align}
In the following, we provide a separate treatment for the fixed-gain and variable-gain relaying schemes.

\subsection{Fixed-Gain Relaying}
For fixed-gain relaying scheme, the relaying gain is given by $w^2 = \frac{P_r}{NP\sigma_1^2+P_I\sigma_I^2+N_0}$, and we have the following key result.
\begin{theorem}\label{theorem:e1}
The outage probability of the $\mbox{N}$-1-1 dual-hop fixed-gain AF relaying systems is given by
\begin{align}
{P}_{\sf out}(\gamma_{th}) &=1-e^{-\frac{{\gamma_{\sf th}}}{\rho_1}}\sum_{m=0}^{N-1}\left(\frac{\gamma_{\sf th}}{\rho_1}\right)^m\frac{2}{m!} \sum_{j=0}^{m}{m\choose j}\sum_{k=0}^{j}{j\choose k}\frac{\Gamma(k+1)\rho_I^k\rho_1^{k+1}}{\left(\rho_I{\gamma_{\sf th}}+\rho_1\right)^{k+1}}\notag\\
&\left(\frac{\gamma_{\sf
th}}{\rho_1}\right)^{\frac{k-j+1}{2}}\left(\frac{N\rho_1+\rho_I+1}{\rho_2}\right)^{\frac{j-k+1}{2}}K_{k-j+1}\left(2\sqrt{\frac{(N\rho_1+\rho_I+1)\gamma_{\sf
th}}{\rho_1\rho_2}}\right),
\end{align}
where $K_v(x)$ is the $v$-th order modified Bessel function of the second kind \cite[Eq. (8.407.1)]{Table}.
\end{theorem}
\proof See Appendix \ref{appendix:theorem:e1}.\endproof

Theorem \ref{theorem:e1} only involves standard functions, and hence offers an efficient means to evaluate the outage
probability of the $\mbox{N}$-1-1 dual-hop fixed-gain AF relaying systems. For the special case $N=1$, the above expression reduces to prior results presented in \cite[Eq. (12)]{H.Sura2} and \cite[Eq. (10)]{Himal12}. To gain more insights, we look into the high
SNR regime, where simple expressions can be obtained.

\begin{theorem}\label{theorem:a1}
At the high SNR regime, i.e., $\rho_2 = \mu \rho_1$, $\rho_1\rightarrow\infty$, the outage probability of the
$\mbox{N}$-1-1 dual-hop fixed-gain AF relaying systems can be approximated as
\begin{align}
P_{\sf out}^{\sf low}(\gamma_{\sf th}) \approx\left\{\begin{array}{cc}
                                                      \left(\frac{1}{\mu}\left(\ln\frac{\mu\rho_1}{\gamma_{\sf th}}+\psi(1)+\psi(2)\right)+\rho_I+1\right)\frac{\gamma_{\sf th}}{\rho_1}, & N=1,\\
                                                       \frac{N}{\mu(N-1)}\frac{\gamma_{\sf th}}{\rho_1}, & N\geq 2,
                                                       \end{array}
                                                       \right.
\end{align}
where $\psi(x)$ is the digamma function \cite[Eq. (8.360.1)]{Table}.
\end{theorem}
\proof See Appendix \ref{appendix:theorem:a1}.\endproof

Theorem \ref{theorem:a1} indicates that fixed-gain relaying schemes only achieve diversity order one regardless of the
number of antennas $N$. However, increasing $N$ helps improve the outage performance by providing extra array gain.
Moreover, Theorem \ref{theorem:a1} suggests that the impact of CCI vanishes at the high SNR regime when $N\geq 2$.

\subsection{Variable-Gain Relaying}
For variable-gain relaying scheme, the relaying gain is given by $w^2 = \frac{P_r}{|\mathbf{h}_1|^2P+|h_I|^2P_I+N_0}$, and we have the following key result.

\begin{theorem}\label{theorem:e2}
The outage probability of the $N$-1-1 variable-gain relaying systems is given by
\begin{multline}
P_{\sf out}(\gamma_{\sf th}) =1-2\frac{e^{-\frac{\gamma_{\sf th}}{\rho_1}-\frac{\gamma_{\sf
th}}{\rho_2}}}{\sigma_I^2}\sum_{m=0}^{N_1-1}\left(\frac{\gamma_{\sf th}}{\rho_1}\right)^m\frac{1}{m!}
\sum_{j=0}^{m}{m\choose j}\left(\frac{1}{\rho_2}\right)^{m-j}\\
\sum_{k=0}^{j}{j\choose k}\left(\frac{\gamma_{\sf th}}{\rho_2}\right)^{j-k}\left(\frac{(\gamma_{\sf th}+1)\gamma_{\sf
th}}{\rho_1\rho_2}\right)^{\frac{k-m+1}{2}}{\cal I}_1(\gamma_{\sf th}),
\end{multline}
where ${\cal I}_1(\gamma_{\sf th})$ is defined as
\begin{align}
{\cal I}_1(\gamma_{\sf th})=\int_0^{\infty}e^{-\left(\frac{P_I\gamma_{\sf
th}}{N_0\rho_1}+\frac{1}{\sigma_I^2}\right)y_3}\left(\frac{y_3P_I}{N_0}+1\right)^{\frac{k+m+1}{2}}K_{k-m+1}\left(2\sqrt{\frac{(\gamma_{\sf
th}+1)\gamma_{\sf th}}{\rho_1\rho_2}\left(\frac{y_3P_I}{N_0}+1\right)}\right)dy_3.
\end{align}
\end{theorem}
\proof See Appendix \ref{appendix:theorem:e2}.\endproof


To the best of the authors' knowledge, the integral ${\cal I}_1$ does not admit a closed-form expression. However, this
single integral expression can be efficiently evaluated numerically, which still provides computational advantage over
the Monte Carlo simulation method. Alternatively, we can use the following closed-form lower bound on the outage
probability, which is tight across the entire SNR range, and becomes exact at the high SNR regime.
\begin{corollary}\label{coro:1}
The outage probability of the $\mbox{N}$-1-1 variable-gain relaying systems is lower bounded by
\begin{align}
P^{\sf low}_{\sf out}(\gamma_{\sf th}) &= 1-e^{-\frac{\gamma_{\sf th}}{\rho_1}-\frac{\gamma_{\sf
th}}{\rho_2}}\sum_{m=0}^{N-1}\left(\frac{\gamma_{\sf th}}{\rho_1}\right)^m\frac{1}{m!}\sum_{j=0}^{m}{m \choose
j}\frac{\Gamma(j+1)\rho_I^j\rho_1^{j+1}}{(\rho_1+\rho_I\gamma_{\sf th})^{j+1}}.
\end{align}
\end{corollary}
\proof We first notice that the end-to-end SINR can be upper bounded by
\begin{align}
\gamma
&=\frac{\frac{|{\bf h}_1|^2P}{N_0}\frac{|h_2|^2P_r}{N_0}}{\left(\frac{|h_3|^2P_I}{N_0}+1\right)\left(\frac{|h_2|^2P_r}{N_0}+1\right)+\frac{|{\bf h}_1|^2P}{N_0}}\leq
\min\left(\frac{\frac{|{\bf h}_1|^2P}{N_0}}{\frac{|h_3|^2P_I}{N_0}+1},\frac{|h_2|^2P_r}{N_0}\right).
\end{align}
Hence, due to the independence of $h_1$, $h_2$ and $h_3$, the outage probability of the system can be lower bounded by
\begin{align}
 P_{\sf out}^{\sf low}(\gamma_{\sf th})
& =1-{\sf Pr}\left(\frac{\frac{|{\bf h}_1|^2P}{N_0}}{\frac{|h_3|^2P_I}{N_0}+1}\geq \gamma_{\sf th}\right){\sf
Pr}\left(\frac{|h_2|^2P_r}{N_0}\geq \gamma_{\sf th}\right).\label{eqn:poutlow}
\end{align}
To this end, the desired result can be computed after some simple algebraic manipulations with the help of Lemma
\ref{lemma:0} presented in Appendix \ref{appendix:lemmas}.
\endproof

%
Now, we look into the high SNR regime, and investigate the diversity order achieved by the system.
\begin{theorem}\label{theorem:a2}
At the high SNR regime, i.e., $\rho_2 = \mu \rho_1$,
$\rho_1\rightarrow\infty$, the outage probability of the $\mbox{N}$-1-1
variable-gain relaying systems can be approximated as
\begin{align}
P_{\sf out}^{\sf low}(\gamma_{\sf th}) \approx\left\{\begin{array}{cc}
                                                      \left(\frac{1}{\mu}+\rho_I+1\right)\frac{\gamma_{\sf th}}{\rho_1}, & N=1,\\
                                                      \frac{1}{\mu} \frac{\gamma_{\sf th}}{\rho_1}, & N\geq 2.
                                                       \end{array}
                                                       \right.
\end{align}
\end{theorem}
\proof See Appendix \ref{appendix:theorem:a2}.\endproof

Clearly, the variable-gain relaying system also achieves diversity order of one. Now, comparing Theorem
\ref{theorem:a2} and Theorem \ref{theorem:a1}, it is evident that the variable-gain relaying scheme outperforms the
fixed-gain relaying scheme at the high SNR regime. Moreover, the performance gain is much more pronounced for small
$N$, and gradually diminishes when $N$ becomes large. Similarly, we see that the impact of CCI disappears when $N\geq
2$, which suggests that implementing multiple antenna at the source can effectively help combat the CCI at the relay.

\section{The 1-1-$\mbox{N}$ System}
This section considers the case where the destination node is equipped with $N$ antennas, while the source and relay
nodes only have a single antenna. Similarly, to conform to the notation convention, we will use scaler $h_1$, $h_I$ and
vector ${\bf h}_2\sim {\cal CN}^{N\times 1}$ to denote the source-relay, interference-relay and relay-destination
links, respectively. After applying the maximum ratio combining at the destination node, the end-to-end SINR can be
expressed as
\begin{align}
\gamma =\frac{|w|^2|{\bf h}_2|^2|{h}_1|^2P}{|w|^2|{\bf h}_2|^2|{h}_I|^2P_I+|w|^2|{\bf h}_2|^2N_0+N_0}.
\end{align}
For notational convenience, we define $y_1 \triangleq |{h}_1|^2$, $y_2 \triangleq |{\bf h}_2|^2$, $y_3 \triangleq |{
h}_I|^2$.

\subsection{Fixed-Gain Relaying}
For fixed-gain relaying scheme, the relaying gain is given by $w^2 =  \frac{P_r}{P\sigma_1^2+P_I\sigma_I^2+N_0}$, and the outage probability of the system is given in the following theorem.

\begin{theorem}\label{theorem:e3}
The outage probability of the 1-1-N fixed-gain relaying systems is given by
\begin{align}
P_{\sf out}(\gamma_{\sf th}) =  1-\frac{2\rho_1e^{-\frac{\gamma_{\sf th}}{\rho_1}}}{\Gamma(N)(\rho_I\gamma_{\sf th}+\rho_1)}\left(\frac{(\rho_1+\rho_I+1)\gamma_{\sf th}}{\rho_1\rho_2}\right)^{\frac{N}{2}}K_N\left(2\sqrt{\frac{(\rho_1+\rho_I+1)\gamma_{\sf th}}{\rho_1\rho_2}}\right).
\end{align}
\end{theorem}
\proof From the definition, the outage probability is given by
\begin{align}
P_{\sf out}(\gamma_{\sf th})= {\sf Pr}\left(y_1\leq \frac{\gamma_{\sf th}}{P}\left(y_3P_I+N_0+\frac{N_0}{|w|^2y_2}\right)\right).
\end{align}
Conditioned on $y_2$ and $y_3$, the outage probability can be shown as
\begin{align}
P_{\sf out}(\gamma_{\sf th})= 1-e^{-\frac{\gamma_{\sf th}N_0}{P\sigma_1^2}}e^{-\frac{\gamma_{\sf th}N_0}{P\sigma_1^2|w|^2y_2}}e^{-\frac{\gamma_{\sf th}y_3P_I}{P\sigma_1^2}}.
\end{align}
Averaging over $y_2$ and $y_3$, the unconditional outage probability can be obtained as
\begin{align}\label{eqn:toutage2}
P_{\sf out}(\gamma_{\sf th})= 1-e^{-\frac{\gamma_{\sf th}N_0}{P\sigma_1^2}}\frac{2}{\Gamma(N)\sigma_2^{2N}}\left(\frac{\gamma_{\sf th}N_0\sigma_2^2}{P\sigma_1^2|w|^2}\right)^{\frac{N}{2}}K_N\left(2\sqrt{\frac{\gamma_{\sf th}N_0}{P\sigma_1^2\sigma_2^2|w|^2}}\right)\frac{1}{\frac{P_I\sigma_I^2\gamma_{\sf th}}{P\sigma_1^2}+1}.
\end{align}
To this end, substituting $w$ into Eq. (\ref{eqn:toutage2}), the desired result can be obtained after some simple algebraic manipulations.
\endproof

Having obtained the exact outage probability expression, we now
establish the asymptotical outage probability approximation at the
high SNR regime.

\begin{theorem}\label{theorem:a3}
At the high SNR regime, i.e., $\rho_2 = \mu \rho_1$, $\rho_1\rightarrow\infty$, the outage probability of the 1-1-N
dual-hop fixed-gain AF relaying systems can be approximated by
\begin{align}
P_{\sf out}(\gamma_{\sf th}) \approx\left(\rho_I+1+\frac{1}{(N-1)\mu}\right)\frac{\gamma_{\sf th}}{\rho_1}, \quad N\geq
2.
\end{align}
\end{theorem}
\proof Utilizing the asymptotic expansion (\ref{eqn:asympt1}), the desired result can be obtained after some basic
algebraic manipulations.\endproof

Theorem \ref{theorem:a3} indicates that the 1-1-N system achieves diversity order one. Also, it suggests that a large
$N$ and relay transmit power helps to reduces the outage probability by providing a larger array gain. Moreover, it shows that the CCI always degrades the outage performance of the system.


\subsection{Variable-Gain Relaying}
For variable-gain relaying scheme, the relaying gain is given by $w^2 = \frac{P_r}{y_1P+y_3P_I+N_0}$, and the outage probability of the system is given in the following theorem.

\begin{theorem}\label{theorem:e4}
The outage probability of the 1-1-N dual-hop variable-gain AF relaying systems can be expressed as
\begin{align}
P_{\sf out}(\gamma_{\sf th}) =1-\frac{2e^{-\frac{\gamma_{\sf th}}{\rho_1}-\frac{\gamma_{\sf
th}}{\rho_2}}}{\sigma_I^2\Gamma(N)}\sum_{k=0}^{N-1}{N-1\choose k}\rho_2^{N-k-1}\left(\frac{(\gamma_{\sf
th}+1)\gamma_{\sf th}}{\rho_1\rho_2}\right)^{\frac{k+1}{2}}{\cal I}_2(\gamma_{\sf th}),
\end{align}
where
\begin{align}
{\cal I}_2(\gamma_{\sf th}) =\int_0^{\infty}e^{-\frac{\gamma_{\sf
th}P_Iy_3}{\rho_1N_0}-\frac{y_3}{\sigma_I^2}}\left(\frac{P_Iy_3}{N_0}+1\right)^{\frac{k+1}{2}}K_{k+1}\left(2\sqrt{\frac{(\gamma_{\sf
th}+1)\gamma_{\sf
th}}{\rho_1\rho_2}\left(\frac{P_Iy_3}{N_0}+1\right)}\right)dy_3.
\end{align}
\end{theorem}
\proof The result can be obtained by following similar lines as in
the proof of Theorem \ref{theorem:e2}, along with some simple
algebraic manipulations.\endproof

\begin{corollary}
The outage probability of the 1-1-N dual-hop variable-gain AF relaying systems is lower bounded by
\begin{align}
P_{\sf out}^{\sf low}(\gamma_{\sf th}) =
1-\frac{\rho_1e^{-\frac{\gamma_{\sf
th}}{\rho_1}}}{\rho_1+\rho_I\gamma_{\sf
th}}\left(1-\frac{1}{\Gamma(N)}\gamma\left(N,\frac{\gamma_{\sf
th}}{\rho_2}\right)\right),
\end{align}
where $\gamma(n,x)$ is the lower incomplete gamma function \cite[Eq. (8.350.1)]{Table}.
\end{corollary}
\proof The result can be obtained by following similar lines as in
the proof of Corollary \ref{coro:1}, along with some simple
algebraic manipulations.\endproof

Now, we look into the high SNR regime, and investigate the diversity order achieved by the system.
\begin{theorem}\label{theorem:a4}
At the high SNR regime, i.e., $\rho_2 = \mu \rho_1$, $\rho_1\rightarrow\infty$, the outage probability of the system
can be approximated as
\begin{align}
P_{\sf out}^{\sf low}(\gamma_{\sf th})
\approx\left(1+\rho_I\right)\frac{\gamma_{\sf th}}{\rho_1}, \quad
N\geq 2.
\end{align}
\end{theorem}
\proof Utilizing the asymptotical expansion of the incomplete gamma function \cite[Eq. (8.354.1)]{Table}, the desired result can be obtained after some simple algebraic manipulations.\endproof

Not surprisingly,  we see that the 1-1-N system with variable-gain relaying also achieves diversity order one. Compared
with Theorem \ref{theorem:a3}, we see that the variable-gain relaying scheme outperforms the fixed-gain relaying scheme
by achieving a higher array gain. Also, Theorem \ref{theorem:a4} suggests a rather interesting result that increasing
$N$ beyond two does not produce any advantage at the high SNR regime.

\section{The 1-N-1 System}
This section considers the case where the relay node is equipped with $N$ antennas, while the source and destination
nodes only have a single antenna. Similarly, to conform to the notation convention, we will use ${\bf h}_1\sim {\cal
CN}^{N\times 1}$, ${\bf h}_I\sim {\cal CN}^{N\times 1}$ and ${\bf h}_2\sim {\cal CN}^{1\times N}$ to denote the
source-relay, interference-relay and relay-destination links, respectively. Then it is easy to show that the end-to-end SINR can be expressed as
\begin{align}\label{eqn:sinrw}
\gamma =\frac{|{\bf h}_2^{\dag}{\bf W}{\bf h}_1|^2P}{|{\bf h}_2{\bf W}{\bf h}_I|^2P_I+|{\bf h}_2^{\dag}{\bf
W}|^2N_0+N_0}.
\end{align}




\subsection{Fixed-Gain Relaying}
With fixed-gain relaying scheme, the relay transformation matrix is simply a scaled identity matrix, i.e., ${\bf W} = w{\bf I}$, with $w^2=\frac{P_r}{NP\sigma_1^2+NP_I\sigma_I^2+N_0}$. Hence, the end-to-end SINR reduces to
\begin{align}
\gamma =\frac{w^2|{\bf h}_2^{\dag}{\bf h}_1|^2P}{w^2|{\bf h}_2^{\dag}{\bf h}_I|^2P_I+w^2|{\bf h}_2^{\dag}|^2N_0+N_0}.
\end{align}

\begin{theorem}\label{theorem:e5}
The outage probability of the 1-N-1 dual-hop fixed-gain AF relaying systems is given by
\begin{align}
P_{\sf out}(\gamma_{\sf th}) =1-\frac{2\rho_1e^{-\frac{\gamma_{\sf th}}{\rho_1}}}{\Gamma(N)(\rho_1+{\rho_I}\gamma_{\sf
th})}\left(\frac{(N\rho_1+N\rho_I+1)\gamma_{\sf
th}}{\rho_1\rho_2}\right)^{\frac{N}{2}}K_N\left(2\sqrt{\frac{(N\rho_1+N\rho_I+1)\gamma_{\sf
th}}{\rho_1\rho_2}}\right).\label{eqn:cdf1}
\end{align}
\end{theorem}
\proof See Appendix \ref{appendix:theorem:e5}.\endproof

\begin{theorem}\label{theorem:a5}
At the high SNR regime, $\rho_2 = \mu\rho_1$, $\rho_1\rightarrow \infty$, the outage probability of the 1-N-1
fixed-gain relaying systems can be approximated as
\begin{align}
P_{\sf out}(\gamma_{\sf th})\approx \left(1+\rho_I+\frac{N}{\mu(N-1)}\right)\frac{\gamma_{\sf th}}{\rho_1}, \quad
N\geq 2.
\end{align}
\end{theorem}
\proof Utilizing the asymptotic expansion (\ref{eqn:asympt1}), the desired result can be obtained after some basic
algebraic manipulations.\endproof

Theorem \ref{theorem:a5} indicates that the 1-N-1 system with fixed-gain AF relaying only achieves diversity order one.
Moreover, it suggests that the CCI degrades the outage performance while increasing $N$ helps improve the outage
performance.

\subsection{Variable-Gain Relaying}
When the channel state information (CSI) is available at the relay node, the optimal relay transformation matrix
$\mathbf{W}$ could be obtained by solving Eq. (\ref{eqn:sinrw}).  However, due to the non-convex nature of the problem,
finding the optimal ${\bf W}$ in analytical form does not seem to be tractable. Therefore, we hereafter propose a
heuristic ${\bf W}$ and investigate its performance.

With CSI at the relay node, it is nature to apply the maximal ratio combining/transmitting principle. Hence, the relay transformation matrix is given by ${\bf W} = w\frac{
{\bf h}_2*{\bf h}_1^{\dag}}{|{\bf h}_2||{\bf h}_1|}$. Depending on the availability of the interference channel information (ICI) at the relay node, we consider to two separate cases.

\subsubsection{Without ICI}
In this case, to meet the power constraint at the relay node, we have
\begin{align}
w^2 = \frac{P_r}{{\tt E}\{{\bf h}_1^{\dag}{\bf h}_1P+|{\bf h}_1^{\dag}{\bf h}_I|^2P_I/|{\bf h}_1|^2+N_0\}} =
\frac{P_r}{NP\sigma_1^2+P_I\sigma_I^2+N_0}.
\end{align}
Hence, the end-to-end SINR can be expressed as
\begin{align}
\gamma =\frac{|{\bf h}_2|^2|{\bf h}_1|^2P}{|{\bf h}_2|^2\frac{|{\bf h}_1{\bf h}_I|^2}{|{\bf h}_1|^2}P_I+|{\bf
h}_2^{\dag}|^2N_0+N_0/w^2}.
\end{align}


\begin{theorem}\label{theorem:e6}
The outage probability of the 1-N-1 dual-hop variable-gain AF relaying systems without ICI can be expressed as
\begin{multline}
P_{\sf out}(\gamma_{\sf th})=1-\frac{2e^{-\frac{\gamma_{\sf th}}{\rho_1}}}{\Gamma(N)}\sum_{m=0}^{N-1}\frac{\gamma_{\sf
th}^m }{\rho_1^{m}m!}\sum_{i=0}^m{m \choose i}\sum_{j=0}^i{i\choose
j}\frac{\rho_I^j\Gamma(j+1)}{\left(\frac{\gamma_{\sf th}\rho_I}{\rho_1}+1\right)^{j+1}}\left(\frac{\gamma_{\sf th}}{\rho_1}\right)^{\frac{N+j-i}{2}}\\
\left(\frac{N\rho_1+\rho_I+1}{\rho_2}\right)^{\frac{N+i-j}{2}}K_{N+j-i}\left(2\sqrt{\frac{(N\rho_1+\rho_I+1)\gamma_{\sf
th}}{\rho_1\rho_2}}\right).
\end{multline}
\end{theorem}
\proof See Appendix \ref{appendix:theorem:e6}.\endproof

\begin{theorem}\label{theorem:a6}
At the high SNR regime, $\rho_2 = \mu\rho_1$, $\rho_1\rightarrow \infty$, the outage probability of the 1-N-1 dual-hop
variable-gain AF relaying systems without ICI can be approximated as
\begin{align}
P_{\sf out}(\gamma_{\sf th}) \approx \sum_{i=0}^{N-1}\frac{(-1)^{N-i}(\ln\frac{N\gamma_{\sf
th}}{\mu\rho_1}-\psi(1)-\psi(N-i+1))}{\Gamma(N)\Gamma(i+1)\Gamma(N-i+1)}\left(\frac{N\gamma_{\sf
th}}{\mu\rho_1}\right)^N.
\end{align}
\end{theorem}
\proof The result can be obtained by following similar lines as in the proof of Theorem \ref{theorem:e1} with the help of Lemma \ref{lemma:00}.
\endproof

\subsubsection{With ICI}
When the ICI is available at the relay node, we have
\begin{align}
w^2 = \frac{P_r}{{\bf h}_1^{\dag}{\bf h}_1P+|{\bf h}_1^{\dag}{\bf h}_I|^2P_I/|{\bf h}_1|^2+N_0}.
\end{align}
Hence, the end-to-end SINR can be expressed as
\begin{align}
\gamma =\frac{|{\bf h}_2|^2|{\bf h}_1|^2P}{|{\bf h}_2|^2\frac{|{\bf h}_1{\bf h}_I|^2}{|{\bf h}_1|^2}P_I+|{\bf
h}_2^{\dag}|^2N_0+\frac{N_0}{P_r}\left(|{\bf h}_1|^2P+\frac{|{\bf h}_1^{\dag}{\bf h}_I|^2}{|{\bf
h}_1|^2}P_I+N_0\right)}.
\end{align}

%

\begin{theorem}\label{theorem:e7}
The outage probability of the 1-N-1 dual-hop variable-gain AF relaying systems with ICI can be expressed as
\begin{multline}
P_{\sf out}(\gamma_{\sf th}) =1-\frac{2e^{-\frac{\gamma_{\sf th}}{\rho_1}-\frac{\gamma_{\sf th}}{\rho_2}}}{\sigma_I^2\Gamma(N)}\sum_{m=0}^{N-1}\left(\frac{\gamma_{\sf th}}{\rho_1}\right)^m\frac{1}{m!}\sum_{j=0}^{m}{m\choose j}\left(\frac{1}{\rho_2}\right)^{m-j}\sum_{k=0}^{N+j-1}{N+j-1\choose k}\\
\left(\frac{\gamma_{\sf th}}{\rho_2}\right)^{N+j-1-k}\left(\frac{(\gamma_{\sf th}+1)\gamma_{\sf th}}{\rho_1\rho_2}\right)^{\frac{k-m+1}{2}}{\cal I}_3(\gamma_{\sf th}),
\end{multline}
where
\begin{align}
{\cal I}_3(\gamma_{\sf th})=\int_0^{\infty}K_{k-m+1}\left(2\sqrt{\frac{(\gamma_{\sf th}+1)\gamma_{\sf
th}}{\rho_1\rho_2}\left(\frac{P_Iy_3}{N_0}+1\right)}\right)\left(\frac{P_Iy_3}{N_0}+1\right)^{\frac{k+m+1}{2}}e^{-\left(\frac{P_I\gamma_{\sf
th}}{\rho_1N_0}+\frac{1}{\sigma_I^2}\right)y_3}dy_3.
\end{align}
\end{theorem}
\proof The result can be obtained by following similar lines as in
the proof of Theorem \ref{theorem:e2}, along with some simple
algebraic manipulations.\endproof

\begin{corollary}\label{coro:3}
The outage probability of the 1-N-1 dual-hop variable-gain AF relaying systems with ICI can be lower bounded by
\begin{align}
P_{\sf out}^{\sf low}(\gamma_{\sf th}) &= 1-\left(1-\frac{\gamma\left(N,\frac{\gamma_{\sf
th}}{\rho_2}\right)}{\Gamma(N)}\right)\left(e^{-\frac{\gamma_{\sf th}}{\rho_1}}\sum_{m=0}^{N-1}\left(\frac{\gamma_{\sf
th}}{\rho_1}\right)^m\frac{1}{m!}\sum_{j=0}^{m}{m \choose
j}\frac{\Gamma(j+1)\rho_I^j\rho_1^{j+1}}{(\rho_1+\rho_I\gamma_{\sf th})^{j+1}}\right).
\end{align}
\end{corollary}
\proof The result can be obtained by following similar lines as in
the proof of Corollary \ref{coro:1}, along with some simple
algebraic manipulations.\endproof

\begin{theorem}\label{theorem:a7}
At the high SNR regime, the outage probability of the system can be approximated
\begin{align}
P_{\sf out}^{\sf low}(\gamma_{\sf th}) \approx
\frac{1}{\Gamma(N+1)}\left({\rho_I}^Ne^{\frac{1}{\rho_I}}\Gamma\left(N+1,\frac{1}{\rho_I}\right)+\frac{1}{\mu^N}\right)\left(\frac{\gamma_{\sf
th}}{\rho_1}\right)^N,\quad N\geq 2,
\end{align}
where $\Gamma(n,x)$ is the upper incomplete gamma function \cite[Eq. (8.350.2)]{Table}.
\end{theorem}
\proof See Appendix \ref{appendix:theorem:a7}.
\endproof

While Theorem \ref{theorem:a5} implies that the fixed-gain relaying scheme only achieves diversity order of one, both
Theorem \ref{theorem:a6} and \ref{theorem:a7} reveal that diversity order of $N$ is achieved by 1-N-1 systems with
variable-gain relaying scheme.

%

\section{Numerical Results and Discussion}
In this section, Monte Carlo simulation results are provided to
validate the analytical expressions presented in the previous
sections. Note, the integral expressions presented in Theorem 3 and
Theorem 7 are evaluated numerically with the build-in functions in
Matlab, i.e., the ``quad'' command, and we choose the default absolute error tolerance value $1.0*10^{-6}$ to control the accuracy of the numerical integration. For all the simulations, we set $\gamma_{\sf th} = 0\mbox{
dB}$, and $\rho_I = 0\mbox{ dB}$. Also, all the simulation results
are obtained by $10^7$ runs. In general, deploying multiple antenna
helps to combat the impact of CCI at the relay node, and the
variable-gain relaying scheme outperforms the fixed-gain relaying
scheme at the high SNR regime, see Table \ref{summary:table} for a
summary of the performance comparison between the two relaying
schemes.

Figure 2 plots the outage probability of the N-1-1 dual-hop AF
relaying systems for both fixed-gain and variable-gain relaying
schemes when $\mu=1$. First of all, we can see that the analytical
results are in exact agreement with the Monte Carlo simulation
results, and the outage lower bound for the variable-gain relaying
system is sufficiently tight across the entire SNR range of
interest, while the high SNR approximations works quite well even at
moderate SNRs (i.e., $\rho_1 = 18\mbox{ dB}$). It can also be
observed that, for both $N = 1$ and $N=2$, the same diversity order
of one is achieved by both the fixed-gain and variable-gain relaying
schemes, which implies that increasing $N$ does not provide
additional diversity gain for the N-1-1 system. However, it does
improve the outage performance of the system by offering extra
coding gain. Moreover, the variable-gain relaying schemes in general
outperforms the fixed-gain relaying schemes.

Figure 3 examines the outage probability of the 1-1-N dual-hop AF
relaying systems for both fixed-gain and variable-gain relaying
schemes when $\mu=1$. Similar to the N-1-1 dual-hop AF relaying
systems, we observe that only diversity order of one is achieved for
both the fixed-gain and variable-gain relaying systems regardless of
$N$, and variable-gain relaying systems achieve superior outage
performance than the fixed-gain relaying systems. However, such
performance gain diminishes gradually $N$ becomes larger. As
illustrated in the figure, the outage gap when $N=10$ is much
narrower when compared with $N=2$. This rather interesting
phenomenon is mainly due to the fact that the outage performance
improvement of the variable-gain relaying schemes due to increasing
$N$ is almost intangible at the high SNR regime, as manifested in
Theorem \ref{theorem:a4}.

Figure 4 illustrates the outage probability of the 1-N-1 dual-hop AF
relaying systems for both fixed-gain and variable-gain relaying
schemes when $N=2$ and $\mu=1$. As shown in the figure, the
diversity order achieved by the fixed-gain relaying scheme is one,
while the diversity order achieved by the variable-gain relaying
schemes is two. Moreover, we see that additional improvement of the
outage performance is achieved when the ICI is available at the
relay node. Both observations suggest the critical importance of
have CSI at the relay node for the 1-N-1 AF dual-hop systems.

Figure 5 provides an outage performance comparison between the
N-1-1, 1-1-N and 1-N-1 dual-hop AF relaying systems with fixed-gain
relaying scheme. Let us first look back at Theorem \ref{theorem:e1},
\ref{theorem:e3} and \ref{theorem:e5}, we see that the coefficients
of the high SNR approximations for the N-1-1, 1-1-N and 1-N-1 are
given by $a_{N11} = \frac{N}{\mu(N-1)}$, $a_{11N} =
\rho_I+1+\frac{1}{\mu(N-1)}$, and $a_{1N1} =
1+\rho_I+\frac{N}{\mu(N-1)}$, respectively. It is easy to observe
that $a_{1N1}\geq \{a_{N11}, a_{11N}\}$. Now the difference of
$a_{N11}$ and $a_{11N}$ can be computed as $a_{N11}-a_{11N} =
\frac{1}{\mu}-(1+\rho_I)$, which suggests the N-1-1 system
outperforms the 1-1-N system only if $\frac{1}{\mu} \leq 1+\rho_I$.
In Figure 4, it can be observed that the 1-N-1 system always has the
worst outage performance, while whether the outage performance of
N-1-1 systems is superior than that of the 1-1-N systems depends on
$\mu$, which confirms the above analysis.

Figure 6 provides an outage performance comparison between the
N-1-1, 1-1-N and 1-N-1 dual-hop AF systems with variable-gain
relaying scheme for $\mu = 0.2, 2$. Recall Theorem \ref{theorem:a6}
and \ref{theorem:a7} that the 1-N-1 system achieves diversity order
of $N$, hence, it definitely outperforms the 1-1-N and 1-N-1 systems
which only achieve diversity order of one. While a close observation
at Theorem \ref{theorem:a2} and \ref{theorem:a4} shows that whether
the 1-1-N system is superior to the N-1-1 system depends on the
relationship between $\frac{1}{\mu}$ and $1+\rho_I$. In Figure 5, we
see that the 1-N-1 system always has the best outage performance,
while the outage performance of the N-1-1 system is better than
1-1-N system when $\mu = 2$, and worse than the 1-1-N when
$\mu=0.2$.

\section{Conclusion}
This paper has investigated the outage performance of dual-hop multiple antenna AF relaying systems with both
fixed-gain and variable-gain relaying schemes. Exact analytical expressions for the outage probability of the systems
under consideration were presented, which provide fast and efficient means to evaluate the outage probability of the
systems. In addition, simple and informative high SNR approximations were derived, while shed lights on how key
parameters, such as CCI, antenna number $N$, and relay power $\mu$, affect the outage performance of the systems.

The findings suggest that, for all three scenarios, the variable-gain relaying scheme outperforms the fixed-gain
relaying scheme. Moreover, for the N-1-1 and 1-1-N systems, the performance advantage of variable-gain relaying scheme
diminishes as $N$ increases. On the contrary, for the 1-N-1 system, the performance advantage of variable-gain relaying
is substantially increased when N becomes large. Moreover, it is demonstrated that whether the N-1-1 system outperforms
the 1-1-N system depends on the interference power and the relay power.

It is explicitly proven that, for all three scenarios, the fixed-gain relaying scheme only achieves diversity order of
one. On the other hand, the variable-gain relaying scheme achieves diversity order of one for the N-1-1 and 1-1-N
systems, but provides diversity order of $N$ for the 1-N-1 system, which suggests that it is beneficial to put the
multiple antennas at the relay node when variable-gain relaying scheme is adopted.

 \appendices
\section{Related Lemmas}\label{appendix:lemmas}
In this section, we present three Lemmas which will be used in the
proof of the main results. Specifically, Lemma 1 will be used in the
derivation of the asymptotical high SNR approximation for the N-1-1
system employing fixed-gain relaying scheme, Lemma 2 will be used in
the derivation of the lower bound of the outage probability of the
N-1-1 system employing variable-gain relaying scheme, while Lemma 3
will be used in the derivation of the exact outage probability of
the N-1-1 system employing variable-gain relaying scheme.
\begin{lemma}\label{lemma:00}
Let $U =y_1\min\left(\frac{1}{y_3+1}, \frac{y_2}{C}\right)$, where $y_1$,  $y_2$ are independent random variables with probability density function (p.d.f.) $f_{y_i}(x) = \frac{x^{N_i-1}}{\Gamma(N_i)}e^{-x}$, $i = 1, 2$. $y_3$ is independently
distributed exponential random variable with p.d.f. $f_{y_3}(x) = \frac{1}{\lambda_3}e^{-\frac{x}{\lambda_3}}$. $C$ is
a positive constant. Then the asymptotical expansion of the cumulative distribution function (c.d.f.) of $U$ near zero can be expressed as
\begin{align}
F_U(x) = \left\{\begin{array}{cc}
                  \frac{Cx}{N_1-1} & N_2 =1\\
                 \sum_{i=0}^{N-1}\frac{(-1)^{N-i}(\ln
Cx-\psi(1)-\psi(N-i+1))}{\Gamma(N)\Gamma(i+1)\Gamma(N-i+1)}(Cx)^N, & N_1 = N_2=N.
                                    \end{array}
                  \right.
\end{align}
\end{lemma}
\proof Define $v\triangleq \left(\frac{1}{y_3+1}, \frac{y_2}{C}\right)$, we first take a look at the asymptotic
behavior of $v$ near zero. To do so, we start by deriving the c.d.f. of $v$ as follows
\begin{align}
F_v(x)& = 1-{\sf Pr}\left(\frac{1}{y_3+1}\geq x\right){\sf Pr}\left(\frac{y_2}{C}\geq x\right)=1-{\sf Pr}\left(y_3\leq \frac{1}{x}-1\right)\left(1-{\sf Pr}\left(y_2\leq Cx\right)\right)\notag\\
& = 1-\left(1-\exp\left(-\frac{1}{\lambda_3}\left(\frac{1}{x}-1\right)\right)\right)(1-F_{y_2}(Cx)).
\end{align}
Hence, conditioned on $y_1$, the c.d.f. of $U$ can be expressed as
\begin{align}
F_U(x)&=1-\left(1-\exp\left(-\frac{y_1}{\lambda_3}\left(\frac{1}{x}-1\right)\right)\right)\left(1-F_{y_2}\left(\frac{Cx}{y_1}\right)\right).
\end{align}
Now, we observe that when $x \rightarrow 0$,
$\exp\left(-\frac{1}{\lambda_3}\left(\frac{y_1}{x}-1\right)\right)\rightarrow 0$, hence, the conditional c.d.f. of $U$
can be well approximated by $F_U(x) \approx F_{y_2}\left(\frac{Cx}{y_1}\right)$.
For the $N_2 = 1$ case, the c.d.f. of $y_2$ near zero can be approximated as $F_{y_2}\left(\frac{Cx}{y_1}\right) \approx \frac{Cx}{y_1}$. To this end, average over $y_1$ yields the desired result.

The $N_1 = N_2 = N$ case is a bit more tricky since we can not approximate the c.d.f. of $y_2$ near zero because of the fact that the resulting integration does not converge. Therefore, we adopt an alternative method. We first explicitly approximate the c.d.f. of $U$ near zero as
\begin{align}\label{eqn:fu}
F_U(x) \approx
1-\frac{2}{\Gamma(N)}\sum_{i=0}^{N-1}\frac{(Cx)^i}{\Gamma(i+1)}(Cx)^{\frac{N-i}{2}}K_{N-i}\left(2\sqrt{Cx}\right).
\end{align}
Invoking the asymptotic expansion of the $K_v(x)$ \cite[Eq. (8.446)]{Table}, we have
\begin{multline}\label{eqn:asympt1}
(Cx)^{\frac{N-i}{2}}K_{N-i}(2\sqrt{Cx})= \frac{1}{2}\sum_{k=0}^{N-i-1}\frac{\Gamma(N-i-k)}{\Gamma(k+1)}\left(-Cx\right)^k\\
-\frac{(-Cx)^{N-i}}{2}\sum_{k=0}^{\infty}\frac{\ln x-\psi(k+1)-\psi(N-i+k+1)}{\Gamma(k+1)\Gamma(N-i+k+1)}(Cx)^k.
\end{multline}
The next key observation is that
\begin{align}
\sum_{i=0}^{N-1}\frac{x^i}{\Gamma(i+1)}\sum_{k=0}^{N-i-1}\frac{\Gamma(N-i-k)}{\Gamma(k+1)}(-x)^k = \Gamma(N).
\end{align}
Hence, the first non-zero term in Eq. (\ref{eqn:fu}) after the asymptotic expansion is given by
\begin{align}
F_U(x) \approx\frac{(Cx)^N}{\Gamma(N)}\sum_{i=0}^{N-1}\frac{(-1)^{N-i}(\ln
Cx-\psi(1)-\psi(N-i+1))}{\Gamma(i+1)\Gamma(N-i+1)},
\end{align}
which completes the proof.
%
%
%

\begin{lemma}\label{lemma:0}
Let $y_1$ and $y_2$ be independent random variables with p.d.f. $f_{y_1}(x) =
\frac{x^{N_1-1}}{\lambda_1^{N_1}\Gamma(N_1)}e^{-\frac{x}{\lambda_1}}$,  $f_{y_2}(x) =
\frac{1}{\lambda_2}e^{-\frac{x}{\lambda_2}}$, and $a$, $b$ are positive constant, then the c.d.f. of random variable $U
\triangleq \frac{ay_1}{by_2+1}$ is given by
\begin{align}
F_U(x) =1-e^{-\frac{x}{a\lambda_1}}\sum_{m=0}^{N_1-1}\left(\frac{x}{a\lambda_1}\right)^m\frac{1}{m!}\sum_{j=0}^{m}{m
\choose j}\frac{\Gamma(j+1)b^j}{\lambda_2\left(\frac{bx}{a\lambda_1}+\frac{1}{\lambda_2}\right)^{j+1}}.
\end{align}
\end{lemma}
\proof Starting from the definition, the c.d.f. of random variable $U$ can be computed as
\begin{align}
F_U(x) &= {\sf Pr}(U\leq x) ={\sf Pr}\left(y_1\leq \frac{x}{a}(by_2+1)\right) = \int_{0}^{\infty}F_{y_1}\left(\frac{x}{a}(bt+1)\right)f_{y_2}(t)dt.
\end{align}
To this end, plunging the corresponding c.d.f. of $y_1$ and p.d.f. of $y_2$, the desired result follows after some
algebraic manipulations.
\endproof
\begin{lemma}\label{lemma:1}
Let $y_1$ and $y_2$ be independent random variables with p.d.f. $f_{y_i}(x) =
\frac{x^{N_i-1}}{\lambda_i^{N_i}\Gamma(N_i)}e^{-\frac{x}{\lambda_i}}$, $i = 1, 2$, and $a$, $b$ are positive constant,
then the c.d.f. of random variable $U \triangleq y_1\frac{y_2-ab}{y_2+a}$ is given by
\begin{multline}
F_U(x)
=1-\frac{e^{-\frac{x}{\lambda_1}-\frac{ab}{\lambda_2}}}{\lambda_2^{N_2}\Gamma(N_2)}\sum_{m=0}^{N_1-1}\left(\frac{x}{\lambda_1}\right)^m\frac{1}{m!}
\sum_{j=0}^{m}{m\choose j}a^{m-j}\sum_{k=0}^{N_2+j-1}{N_2+j-1\choose k}\left(ab\right)^{N_2+j-1-k}\\
2\left(\frac{a(b+1)\lambda_2x}{\lambda_1}\right)^{\frac{k-m+1}{2}}K_{k-m+1}\left(2\sqrt{\frac{a(b+1)x}{\lambda_1\lambda_2}}\right).
\end{multline}
\end{lemma}
\proof Starting from the definition, the c.d.f. of random variable $U$ can be computed as
\begin{align}
F_U(x) &= {\sf Pr}(U\leq x) ={\sf Pr}\left(y_1\frac{y_2-ab}{y_2+a}\leq x\right) = \int_{
ab}^{\infty}F_{y_1}\left(\frac{\left(t+a\right)x}{t-ab}\right)f_{y_2}(t)dt+ \int_0^{ab}f_{y_2}(t)dt.
\end{align}
Utilizing the c.d.f. of random variable $y_1$, the c.d.f. of $U$ can be expressed as
\begin{align}
F_U(x) &=
1-\frac{1}{\lambda_2^{N_2}\Gamma(N_2)}\sum_{m=0}^{N_1-1}\left(\frac{x}{\lambda_1}\right)^m\frac{1}{m!}\int_{ab}^{\infty}e^{-\frac{\left(t+a\right)x}{\left(t-ab\right)\lambda_1}}\left(\frac{t+a}{t-ab}\right)^mt^{N_2-1}e^{-\frac{t}{\lambda_2}}dt
\end{align}
Making a change of variable $s = t-ab$, and simplifying, we have
\begin{align}
F_U(x)
&=1-\frac{e^{-\frac{x}{\lambda_1}-\frac{ab}{\lambda_2}}}{\lambda_2^{N_2}\Gamma(N_2)}\sum_{m=0}^{N_1-1}\left(\frac{x}{\lambda_1}\right)^m\frac{1}{m!}\int_0^{\infty}e^{-\frac{a(b+1)x}{s\lambda_1}-\frac{s}{\lambda_2}}\left(\frac{s+ab+a}{s}\right)^m\left(s+ab\right)^{N_2-1}ds.
\end{align}
Applying the binomial expansion, we arrive at
\begin{align}
F_U(x)
&=1-\frac{e^{-\frac{x}{\lambda_1}-\frac{ab}{\lambda_2}}}{\lambda_2^{N_2}\Gamma(N_2)}\sum_{m=0}^{N_1-1}\left(\frac{x}{\lambda_1}\right)^m\frac{1}{m!}\sum_{j=0}^{m}{m\choose j}a^{m-j}\int_0^{\infty}e^{-\frac{a(b+1)x}{s\lambda_1}-\frac{s}{\lambda_2}}\left(s+ab\right)^{N_2+j-1}s^{-m}ds\notag\\
&=1-\frac{e^{-\frac{x}{\lambda_1}-\frac{ab}{\lambda_2}}}{\lambda_2^{N_2}\Gamma(N_2)}\sum_{m=0}^{N_1-1}\left(\frac{x}{\lambda_1}\right)^m\frac{1}{m!}\sum_{j=0}^{m}{m\choose
j}a^{m-j}\sum_{k=0}^{N_2+j-1}{N_2+j-1\choose
k}\left(ab\right)^{N_2+j-1-k}\notag\\
&\int_0^{\infty}e^{-\frac{a(b+1)x}{s\lambda_1}-\frac{s}{\lambda_2}}s^{k-m}ds.
\end{align}
To this end, the desired result can be obtained with the help of \cite[Eq. (3.471.9)]{Table}.
\endproof

\section{Proof for the $N$-1-1 Systems}
\subsection{Proof of Theorem \ref{theorem:e1}}\label{appendix:theorem:e1}
Starting from the definition, the outage probability can be expressed as
\begin{align}
{P}_{\sf out}(\gamma_{\sf th}) = {\sf Pr}\left(\frac{y_1P}{y_3P_I+N_0+N_0/(|w|^2y_2)}\leq \gamma_{\sf th}\right),
\end{align}
where $y_1 \triangleq |{\bf h}_1|^2$, $y_2 \triangleq |h_2|^2$, $y_3 \triangleq |{ h}_I|^2$.

Conditioned on $y_2$ and $y_3$, the outage probability can be evaluated as
\begin{align}
&{P}_{\sf out}(\gamma_{\sf th}) = 1-e^{-\frac{y_3P_I{\gamma_{\sf th}}}{\sigma_1^2P}}e^{-\frac{N_0{\gamma_{\sf th}}}{\sigma_1^2P}}e^{-\frac{N_0{\gamma_{\sf th}}}{\sigma_1^2P|w|^2y_2}}\sum_{m=0}^{N-1}\left(\frac{\gamma_{\sf th}N_0}{P}\right)^m\frac{\left(\frac{y_3P_I}{N_0}+\frac{1}{|w|^2y_2}+1\right)^m}{\sigma_1^{2m}m!} \notag\\
&=1-e^{-\frac{y_3P_I{\gamma_{\sf th}}}{\sigma_1^2P}}e^{-\frac{N_0{\gamma_{\sf th}}}{\sigma_1^2P}}e^{-\frac{N_0{\gamma_{\sf th}}}{\sigma_1^2P|w|^2y_2}}\sum_{m=0}^{N-1}\left(\frac{\gamma_{\sf th}N_0}{P\sigma_1^{2}}\right)^m\frac{1}{m!} \sum_{j=0}^{m}{m\choose j}\left(\frac{y_3P_I}{N_0}+\frac{1}{|w|^2y_2}\right)^j\notag\\
&=1-e^{-\frac{y_3P_I{\gamma_{\sf th}}}{\sigma_1^2P}}e^{-\frac{N_0{\gamma_{\sf
th}}}{\sigma_1^2P}}e^{-\frac{N_0{\gamma_{\sf th}}}{\sigma_1^2P|w|^2y_2}}\sum_{m=0}^{N-1}\left(\frac{\gamma_{\sf
th}N_0}{P\sigma_1^{2}}\right)^m\frac{1}{m!} \sum_{j=0}^{m}{m\choose j}\sum_{k=0}^{j}{j\choose
k}\left(\frac{y_3P_I}{N_0}\right)^k\left(\frac{1}{|w|^2y_2}\right)^{j-k}.
\end{align}
Hence, averaging over $y_2$ and $y_3$, the unconditional outage probability can be
computed as
\begin{align}\label{eqn:tempout1}
{P}_{\sf out}(\gamma_{\sf th})&=1-e^{-\frac{N_0{\gamma_{\sf th}}}{\sigma_1^2P}}\sum_{m=0}^{N-1}\left(\frac{\gamma_{\sf th}N_0}{P\sigma_1^{2}}\right)^m\frac{1}{m!} \sum_{j=0}^{m}{m\choose j}\sum_{k=0}^{j}{j\choose k}\left(\frac{P_I}{N_0}\right)^k\left(\frac{P_I{\gamma_{\sf th}}}{\sigma_1^2P}+\frac{1}{\sigma_I^2}\right)^{-k-1}\notag\\
&\frac{\Gamma(k+1)}{\sigma_I^2}\left(\frac{1}{|w|^2}\right)^{j-k}\frac{2}{\sigma_2^2}\left(\frac{N_0\gamma_{\sf
th}\sigma_2^2}{\sigma_1^2P|w|^2}\right)^{\frac{k-j+1}{2}}K_{k-j+1}\left(2\sqrt{\frac{N_0\gamma_{\sf
th}}{P|w|^2\sigma_1^2\sigma_2^2}}\right).
\end{align}
To this end, substituting $w$ into Eq. (\ref{eqn:tempout1}), the desired result can be obtained after some simple
algebraic manipulations.

\subsection{Proof of Theorem \ref{theorem:a1}}\label{appendix:theorem:a1}
We find it convenient to give a separate treatment for the $N=1$ and $N\geq 2$ cases. When $N=1$, the outage
probability reduce to
\begin{align}
P_{\sf out} (\gamma_{\sf th}) = 1 - 2e^{-\frac{\gamma_{\sf th}}{\rho_1}}\sqrt{\frac{\gamma_{\sf
th}}{\mu\rho_1}}\left(1+\frac{\rho_I\gamma_{\sf th}}{\rho_1}\right)^{-1}K_1\left(2\sqrt{\frac{\gamma_{\sf
th}}{\mu\rho_1}}\right).
 \end{align}
Applying the asymptotical expansion of $K_v(x)$ according to Eq. (\ref{eqn:asympt1}), we have
 \begin{align}
P_{\sf out}(\gamma_{\sf th}) \approx 1-\left(1-\frac{\gamma_{\sf th}}{\rho_1}\right)\left(1-\frac{\rho_I\gamma_{\sf
th}}{\rho_1}\right)\left(1+\frac{\gamma_{\sf th}}{\mu\rho_1}\left(\ln\frac{\gamma_{\sf
th}}{\mu\rho_1}-\psi(1)-\psi(2)\right)\right)
 \end{align}
To this end, the desired result can be obtained after some simple algebraic manipulations.

When $N\geq 2$, due to the complex multi-summation of $K_v(x)$ in the outage expression, directly utilizing the asymptotic expansion Eq. (\ref{eqn:asympt1}) does not seem to be tractable. Therefore, we adopt the following alternative approach. We note that the end-to-end SINR is statistically equivalent to
\begin{align}
\bar{\gamma} =\frac{\bar{y}_1\bar{y}_2\rho_1\rho_2}{(\bar{y}_3\rho_I+1)\bar{y}_2\rho_2+(1+N\rho_1+\rho_I)},
\end{align}
where $\bar{y}_i$ has the p.d.f. $f_{\bar{y}_i}(x) = \frac{x^{N_i-1}}{\Gamma(N_i)}e^{-x}$, and $N_1 = N$, $N_2 = N_3
=1$. Hence, the outage probability of the system can be alternatively computed as
\begin{align}
{P}_{\sf out}(\gamma_{\sf th}) ={\sf
Pr}\left(\frac{\bar{y}_1\bar{y}_2}{(\bar{y}_3\rho_I+1)\bar{y}_2+(1+N\rho_1+\rho_I)/\rho_2}\leq \frac{\gamma_{\sf
th}}{\rho_1}\right).
\end{align}
At the high SNR regime, the outage probability can be tightly lower bounded by
\begin{align}
{P}_{\sf out}(\gamma_{\sf th}) \geq{\sf
Pr}\left(\min\left(\frac{\bar{y}_1\bar{y}_2}{N/\mu},\frac{\bar{y}_1}{\bar{y}_3\rho_I+1}\right)\leq \frac{\gamma_{\sf
th}}{\rho_1}\right).
\end{align}
To this end, involving Lemma \ref{lemma:00} yields the desired result.

\subsection{Proof of Theorem \ref{theorem:e2}}\label{appendix:theorem:e2}
The outage probability of the system can be expressed as
\begin{align}
P_{\sf out}(\gamma_{\sf th}) &= {\sf Pr}\left(\frac{y_1y_2PP_r}{(y_2P_r+N_0)(y_3P_I+N_0)+y_1PN_0}\leq \gamma_{\sf th}\right)\notag\\
&= {\sf Pr}\left(y_1\frac{y_2-\frac{N_0\gamma_{\sf th}}{P_r}}{y_2+\frac{N_0}{P_r}}\leq \frac{\gamma_{\sf
th}}{P}(y_3P_I+N_0)\right).
\end{align}
Invoking Lemma \ref{lemma:1}, we obtain the following outage probability expression conditioned on $y_3$
\begin{align}
&P_{\sf out}(\gamma_{\sf th}) =1-\frac{e^{-\frac{\gamma_{\sf th}}{P\sigma_1^2}(y_3P_I+N_0)-\frac{N_0\gamma_{\sf
th}}{P_r\sigma_2^2}}}{\sigma_2^{2}}\sum_{m=0}^{N_1-1}\left(\frac{\gamma_{\sf th}N_0}{P\sigma_1^2}\right)^m\frac{1}{m!}
\sum_{j=0}^{m}{m\choose j}\left(\frac{N_0}{P_r}\right)^{m-j}\sum_{k=0}^{j}{j\choose k}\left(\frac{N_0\gamma_{\sf th}}{P_r}\right)^{j-k}\notag\\
&2\left(\frac{N_0^2(\gamma_{\sf th}+1)\gamma_{\sf
th}\sigma_2^2}{P_rP\sigma_1^2}\right)^{\frac{k-m+1}{2}}\left(\frac{y_3P_I}{N_0}+1\right)^{\frac{k+m+1}{2}}K_{k-m+1}\left(2\sqrt{\frac{N_0^2(\gamma_{\sf
th}+1)\gamma_{\sf th}}{PP_r\sigma_1^2\sigma_2^2}\left(\frac{y_3P_I}{N_0}+1\right)}\right).
\end{align}
To this end, the desired result can be obtained by further averaging over $y_3$, along with some simple basic algebraic manipulations.


\subsection{Proof of Theorem \ref{theorem:a2}}\label{appendix:theorem:a2}
Starting from Eq. (\ref{eqn:poutlow}), conditioned on $y_3$, the outage lower bound can be expressed as
\begin{align}
&P_{\sf out}^{\sf low}(\gamma_{\sf th}) =1-\left(1-F_{y_1}\left(\frac{\gamma_{\sf
th}N_0}{P}\left(\frac{y_3P_I}{N_0}+1\right)\right)\right)\left(1-F_{y_2}\left(\frac{\gamma_{\sf
th}N_0}{P_r}\right)\right)\notag\\
&=F_{y_1}\left(\frac{\gamma_{\sf th}N_0}{P}\left(\frac{y_3P_I}{N_0}+1\right)\right)+F_{y_2}\left(\frac{\gamma_{\sf
th}N_0}{P_r}\right)-F_{y_2}\left(\frac{\gamma_{\sf th}N_0}{P_r}\right)F_{y_1}\left(\frac{\gamma_{\sf
th}N_0}{P}\left(\frac{y_3P_I}{N_0}+1\right)\right)\label{eqn:outlow1}.
\end{align}
When $\rho_1$ becomes large, utilizing the asymptotic expansion of lower incomplete gamma function \cite[Eq. (8.354.1)]{Table}, it is easy to show that
\begin{align}
F_{y_1}\left(\frac{\gamma_{\sf
th}N_0}{P}\left(\frac{y_3P_I}{N_0}+1\right)\right)\approx\frac{\left(\frac{y_3P_I}{N_0}+1\right)^N}{\Gamma(N+1)}\left(\frac{\gamma_{\sf
th}}{\rho_1}\right)^N,
\end{align}
and
\begin{align}
F_{y_2}\left(\frac{\gamma_{\sf th}N_0}{P_r}\right) \approx \frac{\gamma_{\sf th}}{\mu\rho_1}.
\end{align}
Hence, quick observation reveals that the outage probability is dominated by the second term in Eq. (\ref{eqn:outlow1})
when $N\geq 2$. On the other hand, when $N=1$, the outage probability can be approximated by
\begin{align}
P_{\sf out}^{\sf low}(\gamma_{\sf th}) \approx \left(\frac{{\tt E}
\{y_3\}P_I}{N_0}+\frac{1}{\mu}+1\right)\frac{\gamma_{\sf th}}{\rho_1}.
\end{align}
Thus, the desired result follows after explicitly computing the first moment of $y_3$.

\section{Proof for the 1-$N$-1 systems}
\subsection{Proof of Theorem \ref{theorem:e5}}\label{appendix:theorem:e5}
We start the proof by expressing the end-to-end SINR as
\begin{align}
\gamma =\frac{Py_1}{P_Iy_2+N_0+\frac{N_0}{a^2y_3}},
\end{align}
where  $y_1\triangleq\frac{|{\bf h}_2^{\dag}{\bf h}_1|^2}{|{\bf h}_2|^2}$,
$y_2\triangleq\frac{|{\bf h}_2^{\dag}{\bf h}_I|^2}{|{\bf h}_2|^2}$ and $y_3\triangleq |{\bf h}_2|^2$.

From \cite{J.Cui}, we
know that $y_1$ and $y_2$ follows exponential distribution with parameter $\sigma_1^2$ and $\sigma_I^2$, respectively.
Moreover, $y_1$, $y_2$, and $y_3$ are mutually independent. Hence, conditioned on $y_2$ and $y_3$, the outage probability of the system can be computed as
\begin{align}
P_{\sf out}(\gamma_{\sf th}) & = 1-e^{-\frac{N_0\gamma_{\sf th}}{P\sigma_1^2}}e^{-\frac{P_I\gamma_{\sf
th}y_2}{P\sigma_1^2}}e^{-\frac{N_0\gamma_{\sf th}}{a^2Py_3\sigma_1^2}}.
\end{align}
To this end, averaging over $y_2$ and $y_3$, we have
\begin{align}
P_{\sf out}(\gamma_{\sf th}) &=
1-e^{-\frac{N_0\gamma_{\sf th}}{P\sigma_1^2}}\int_0^{\infty}e^{-\frac{P_I\gamma_{\sf th}y_2}{P\sigma_1^2}}\frac{1}{\sigma_I^2}e^{-\frac{y_2}{\sigma_I^2}}dy_2\int_0^{\infty}e^{-\frac{N_0x}{a^2P\sigma_1^2y_3}}\frac{1}{\sigma_2^{2N}\Gamma(N)}y_3^{N-1}e^{-\frac{y_3}{\sigma_2^2}}dy_3\notag\\
&=1-\frac{2}{\sigma_2^{2N}\Gamma(N)}\frac{e^{-\frac{N_0\gamma_{\sf th}}{P\sigma_1^2}}}{1+\frac{P_I\gamma_{\sf
th}\sigma_I^2}{P\sigma_1^2}}\left(\frac{N_0\sigma_2^2\gamma_{\sf
th}}{a^2P\sigma_1^2}\right)^{\frac{N}{2}}K_N\left(2\sqrt{\frac{N_0\gamma_{\sf
th}}{a^2P\sigma_1^2\sigma_2^2}}\right).\label{eqn1:cdf1}
\end{align}
Finally, substituting $w$ into Eq. (\ref{eqn1:cdf1}), the desired result follows after some simple algebraic
manipulations.

\subsection{Proof of Theorem \ref{theorem:e6}}\label{appendix:theorem:e6}
The end-to-end SINR can be alternatively expressed as
\begin{align}
\gamma =\frac{y_1y_2P}{y_2y_3P_I+y_2N_0+\frac{N_0}{w^2}},
\end{align}
where $y_1 \triangleq |{\bf h}_1|^2$ and $y_2\triangleq |{\bf h}_2|^2$, and $y_3 \triangleq \frac{|{\bf
h}_1^{\dag}{\bf h}_||^2}{|{\bf h}_1|^2}$. Noticing that $y_3$ is exponentially distributed with parameter $\sigma_I^2$, and $y_3$ is independent of $y_1$, the outage probability conditioned on $y_2$ and $y_3$ can be computed as
\begin{align}
P_{\sf out}(\gamma_{\sf th})&=1-e^{-\frac{\gamma_{\sf th}N_0}{P\sigma_1^2}}e^{-\frac{\gamma_{\sf th}P_I}{P\sigma_1^2}y_3}e^{-\frac{\gamma_{\sf th}N_0}{Pw^2\sigma_1^2y_2}}\sum_{m=0}^{N-1}\frac{\gamma_{\sf th}^mN_0^m}{\sigma_1^{2m}m!P^m}\left(\frac{P_I}{N_0}y_3+\frac{1}{w^2y_2}+1\right)^m.
\end{align}
Now, applying the binomial expansion, we have
\begin{align}
P_{\sf out}(\gamma_{\sf th})&=1-e^{-\frac{\gamma_{\sf th}N_0}{P\sigma_1^2}}e^{-\frac{\gamma_{\sf th}P_I}{P\sigma_1^2}y_3}e^{-\frac{\gamma_{\sf th}N_0}{Pw^2\sigma_1^2y_2}}\sum_{m=0}^{N-1}\frac{\gamma_{\sf th}^mN_0^m}{\sigma_1^{2m}m!P^m}\sum_{i=0}^{m}{m\choose
i}\sum_{j=0}^{i}{i\choose j}\left(\frac{P_Iy_3}{N_0}\right)^j\left(\frac{1}{w^2y_2}\right)^{i-j}.\notag
\end{align}
Averaging over $y_3$ and $y_2$, we have
\begin{align}\label{eqn:outage11}
P_{\sf out}(\gamma_{\sf th})&=1-\frac{2e^{-\frac{\gamma_{\sf th}N_0}{P\sigma_1^2}}}{\sigma_2^{2N}\Gamma(N)}\sum_{m=0}^{N-1}\frac{\gamma_{\sf th}^mN_0^m}{\sigma_1^{2m}m!P^m}\sum_{i=0}^m{m
\choose i}\sum_{j=0}^i{i\choose
j}\left(\frac{P_I}{N_0}\right)^j\left(\frac{1}{w^2}\right)^{i-j}\frac{1}{\sigma_I^2}\frac{\Gamma(j+1)}{\left(\frac{\gamma_{\sf th}P_I}{P\sigma_1^2}+\frac{1}{\sigma_I^2}\right)^{j+1}}\notag\\
&\left(\frac{\gamma_{\sf th}N_0\sigma_2^2}{Pw^2\sigma_1^2}\right)^{\frac{N+j-i}{2}}K_{N+j-i}\left(2\sqrt{\frac{\gamma_{\sf th}N_0}{Pw^2\sigma_1^2\sigma_2^2}}\right).
\end{align}
Finally, substituting $w$ into Eq. (\ref{eqn:outage11}), the desired result follows after some simple algebraic
manipulations.

\subsection{Proof of Theorem \ref{theorem:a7}}\label{appendix:theorem:a7}
Due to the double summation involved in Corollary \ref{coro:3}, it is difficult to obtain the asymptotic expansion directly. Hence, we adopt a different approach. Following similar lines as in the proof of Corollary \ref{coro:1}, the outage lower bound can be expressed as
\begin{align}
 P_{\sf out}^{\sf low}(\gamma_{\sf th})
& =1-{\sf Pr}\left(\frac{\frac{y_1P}{N_0}}{\frac{y_3P_I}{N_0}+1}\geq \gamma_{\sf th}\right){\sf
Pr}\left(\frac{y_2P_r}{N_0}\geq \gamma_{\sf th}\right),\label{eqn:poutlow1}
\end{align}
where $y_1 \triangleq |{\bf h}_1|^2$ and $y_2\triangleq |{\bf h}_2|^2$, and $y_3 \triangleq \frac{|{\bf h}_1^{\dag}{\bf h}_||^2}{|{\bf h}_1|^2}$.

Conditioned on $y_3$, the outage lower bound can be expressed as
\begin{align}
P_{\sf out}^{\sf low}(\gamma_{\sf th})
& =1-\left(1-\frac{\gamma\left(N,\frac{\gamma_{\sf th}}{\rho_2}\right)}{\Gamma(N)}\right)\left(1-\frac{\gamma\left(N,\left(\frac{y_3P_I}{N_0}+1\right)\frac{\gamma_{\sf th}}{\rho_1}\right)}{\Gamma(N)}\right).
\end{align}
Then, utilizing the asymptotic expansion of incomplete gamma function \cite[Eq. (8.354.1)]{Table},  the outage lower bound can be approximated as
\begin{align}
P_{\sf out}^{\sf low}(\gamma_{\sf th})\approx \frac{1}{\Gamma(N+1)}\left(\frac{1}{\mu^N}+\left(\frac{y_3P_I}{N_0}+1\right)^N\right)\left(\frac{\gamma_{\sf th}}{\rho_1}\right)^N.
\end{align}
Finally, averaging over $y_3$ yields the desired result.

\newpage
\begin{figure}
\centering
\includegraphics[scale=1]{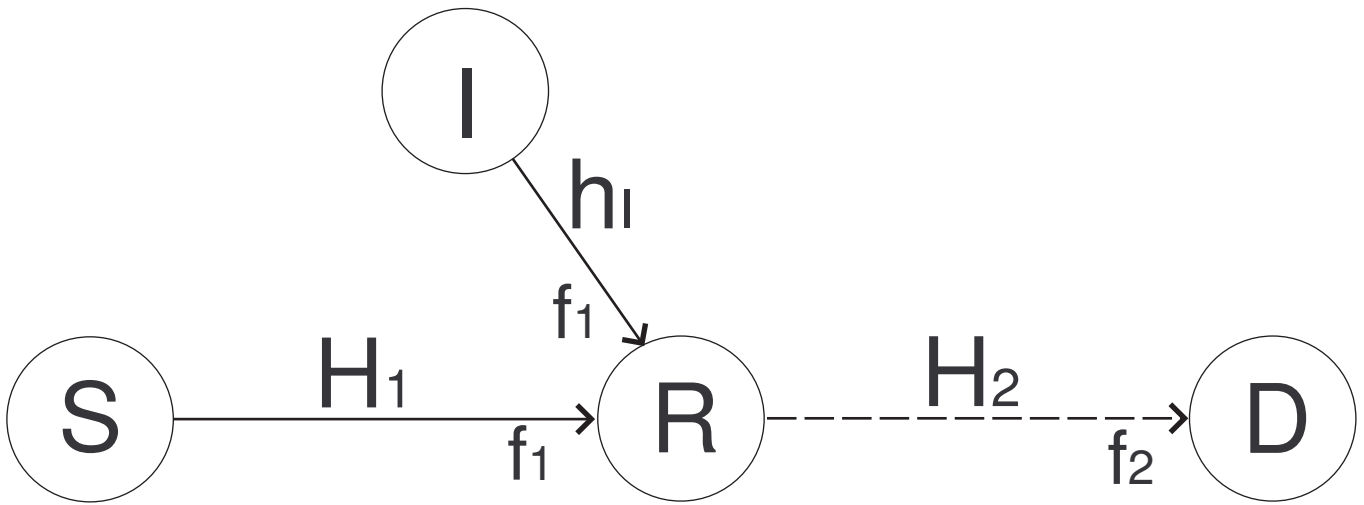}
\caption{System model: S, R, D and I denote source, relay, destination and interferer node, respectively. The source-relay and relay-destination link operate over different frequency $\mbox{f}_1$ and $\mathbf{f}_2$. The signal at R is corrupted by a single dominate interference I and AWGN, while the signal at D is degraded by AWGN only.}\label{fig:fig0}
\end{figure}

\begin{table}
\centering
\caption{High SNR Outage Performance Comparison}
\begin{tabular}{|c|c|p{3.8cm}|p{3.8cm}|p{3.8cm}|}
\hline
\multicolumn{2}{|c|}{}& N-1-1 &1-1-N & 1-N-1\\
\hline
\multirow{3}{*}{Fixed-gain} &Key results& Theorem 2  & Theorem 6  &Theorem 10\\
\cline{2-5}
&Diversity order& 1 & 1& 1\\
\cline{2-5}
&Impact of N& Increasing N provides some array gain, but such improvement diminishes as N becomes larger. & Increasing N provides some array gain, but such improvement diminishes as N becomes larger&Increasing N provides some array gain, but such improvement diminishes as N becomes larger\\
\hline
\multirow{3}{*}{Variable-gain} &Key results& Theorem 4 &Theorem 8 &Theorem 12 \& 14\\\cline{2-5}
&Diversity order&  1& 1& N\\\cline{2-5}
&Impact of N&Same outage performance for $\mbox{N}\geq 2$ & Same outage performance for $\mbox{N}\geq 2$&Increasing N helps to achieve higher diversity order\\
\hline
\end{tabular}
\label{summary:table}
\end{table}
\begin{figure}
\centering
\includegraphics[scale=1.3]{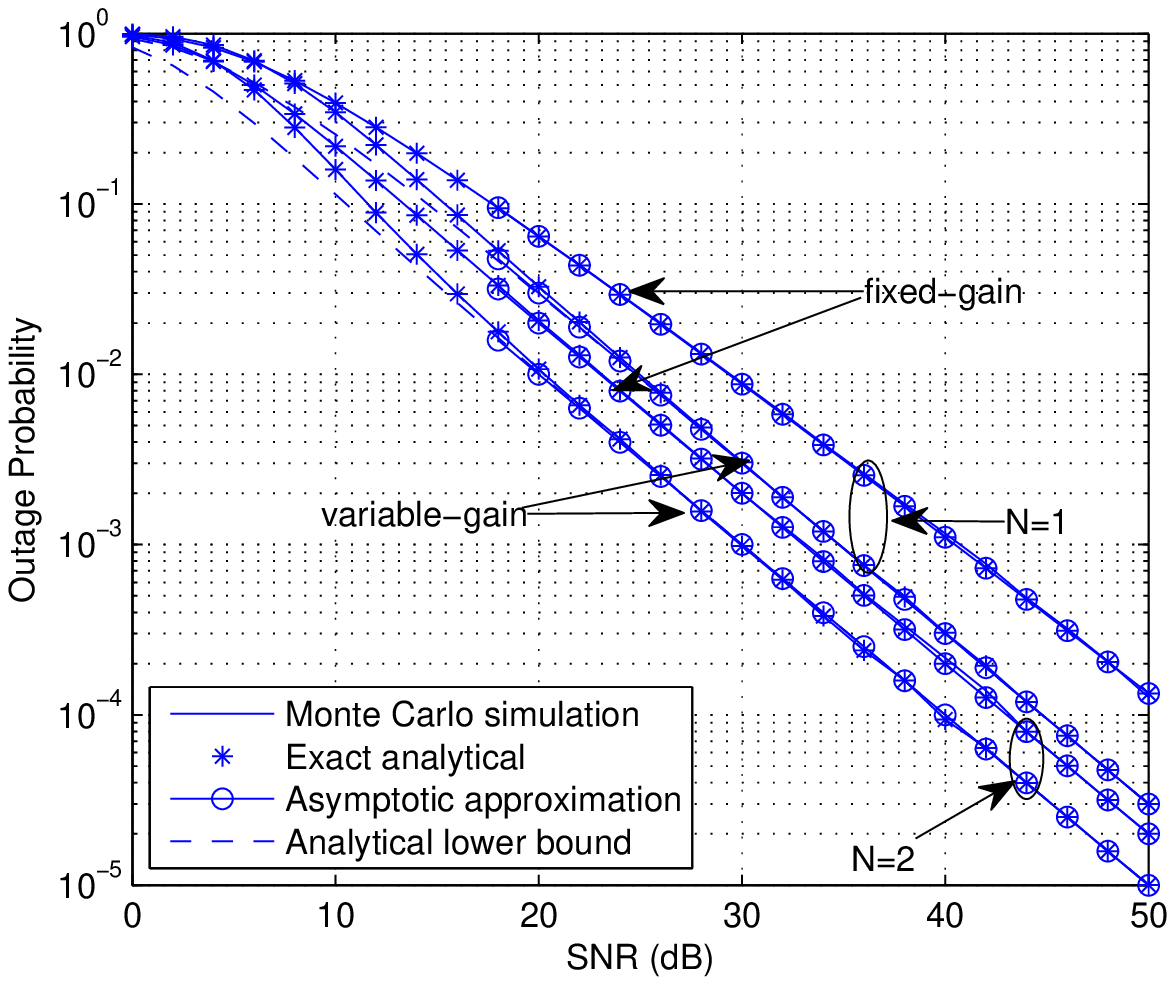}
\caption{Outage probability of the N-1-1 dual-hop relaying systems: fixed-gain vs. variable-gain.}\label{fig:fig1}
\end{figure}

\begin{figure}
\centering
\includegraphics[scale=1.3]{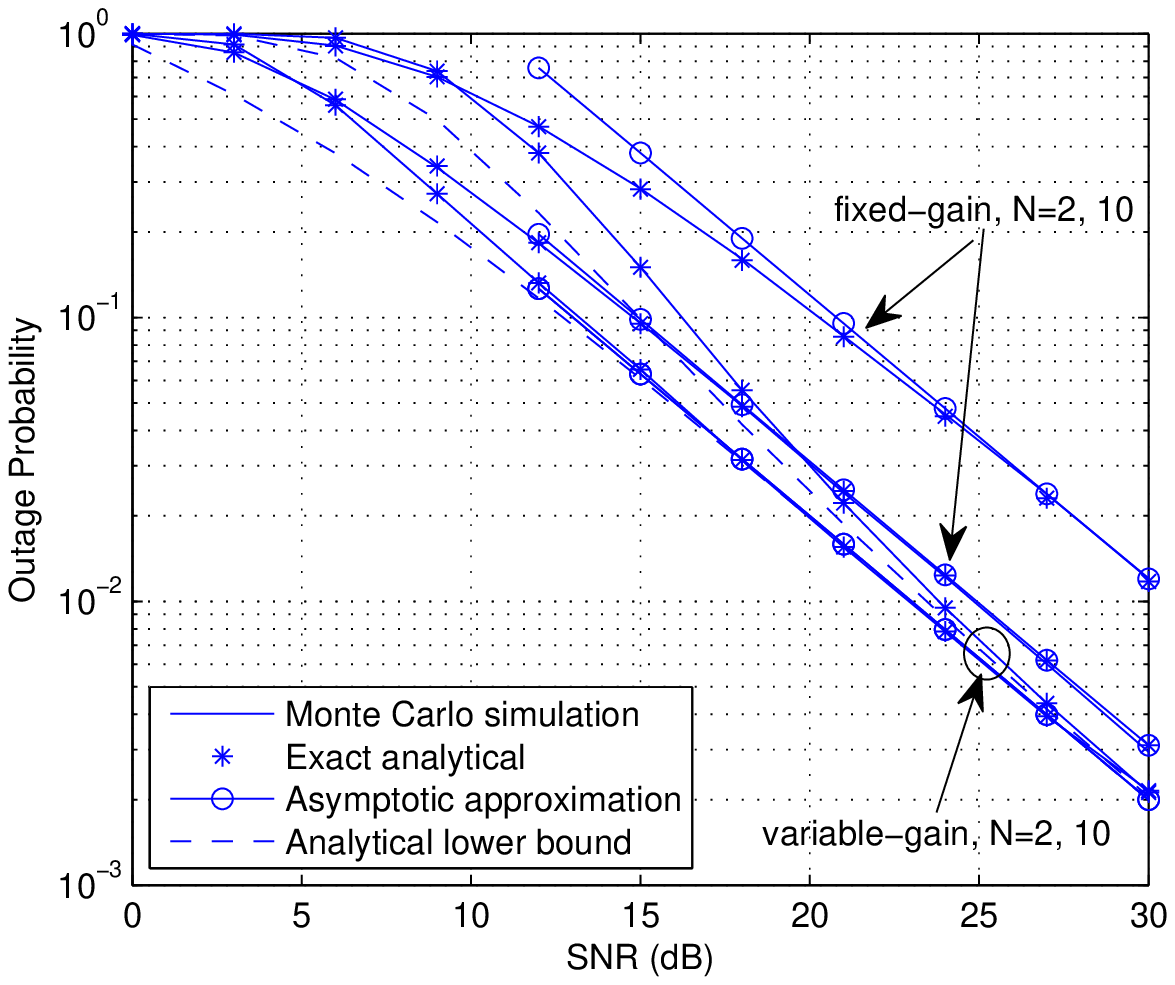}
\caption{Outage probability of the 1-1-N dual-hop relaying systems with different $N$: fixed-gain vs.
variable-gain.}\label{fig:fig2}
\end{figure}

\begin{figure}
\centering
\includegraphics[scale=1.3]{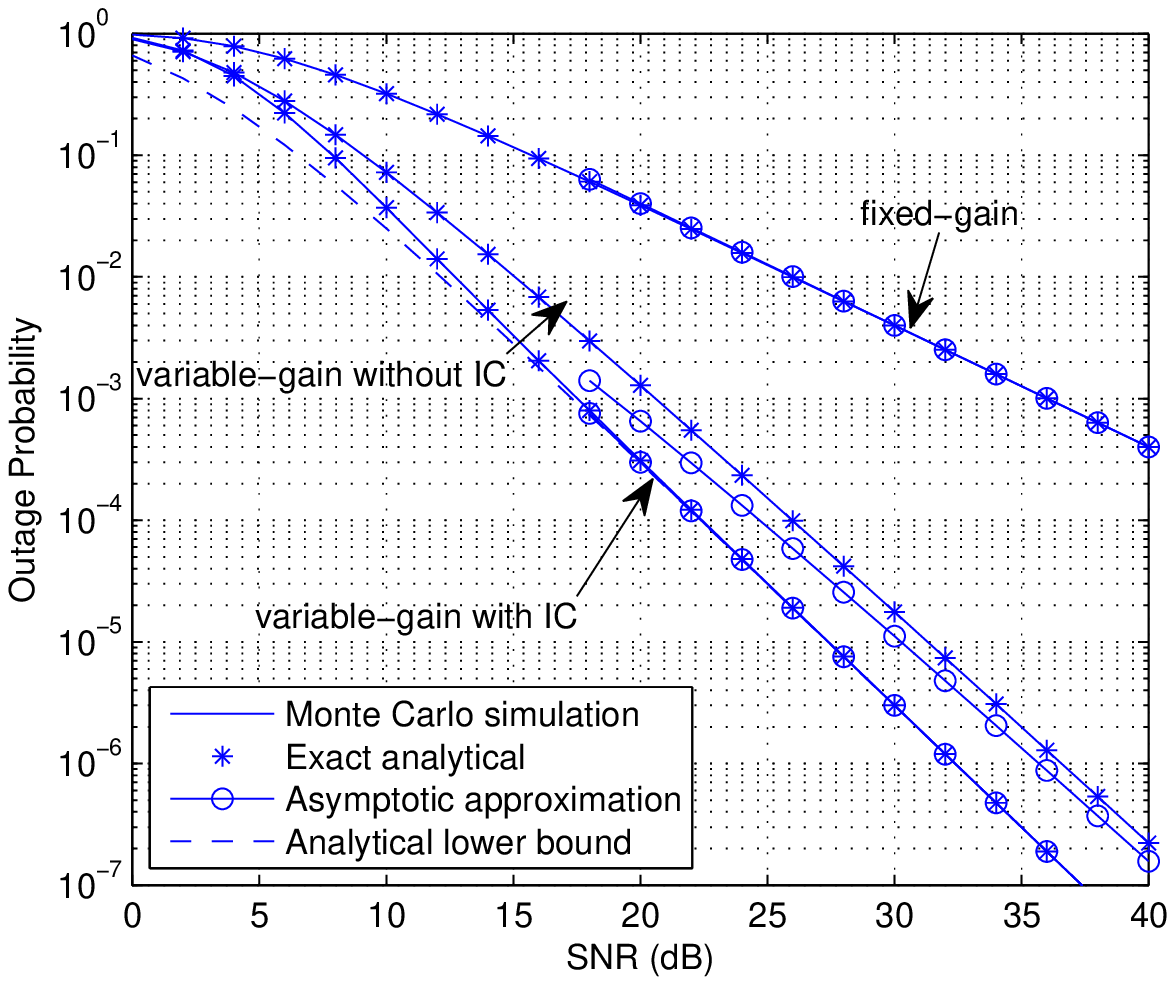}
\caption{Outage probability of the 1-N-1 dual-hop relaying systems with $N=2$: fixed-gain vs.
variable-gain.}\label{fig:fig3}
\end{figure}

\begin{figure}
\centering
\includegraphics[scale=1.3]{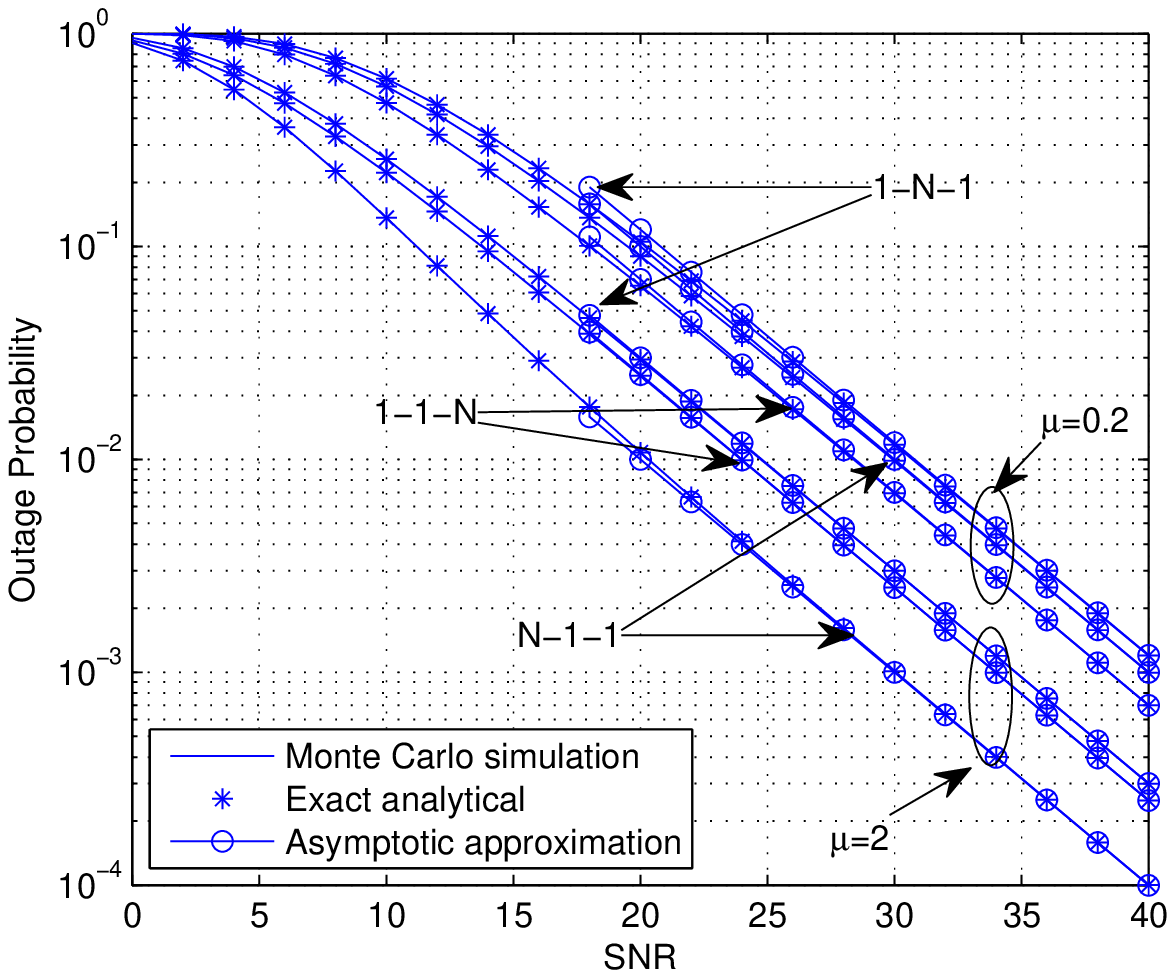}
\caption{Comparison of the outage probability of N-1-1, 1-1-N and 1-N-1 dual-hop systems with fixed-gain
relaying.}\label{fig:fig4}
\end{figure}

\begin{figure}
\centering
\includegraphics[scale=1.3]{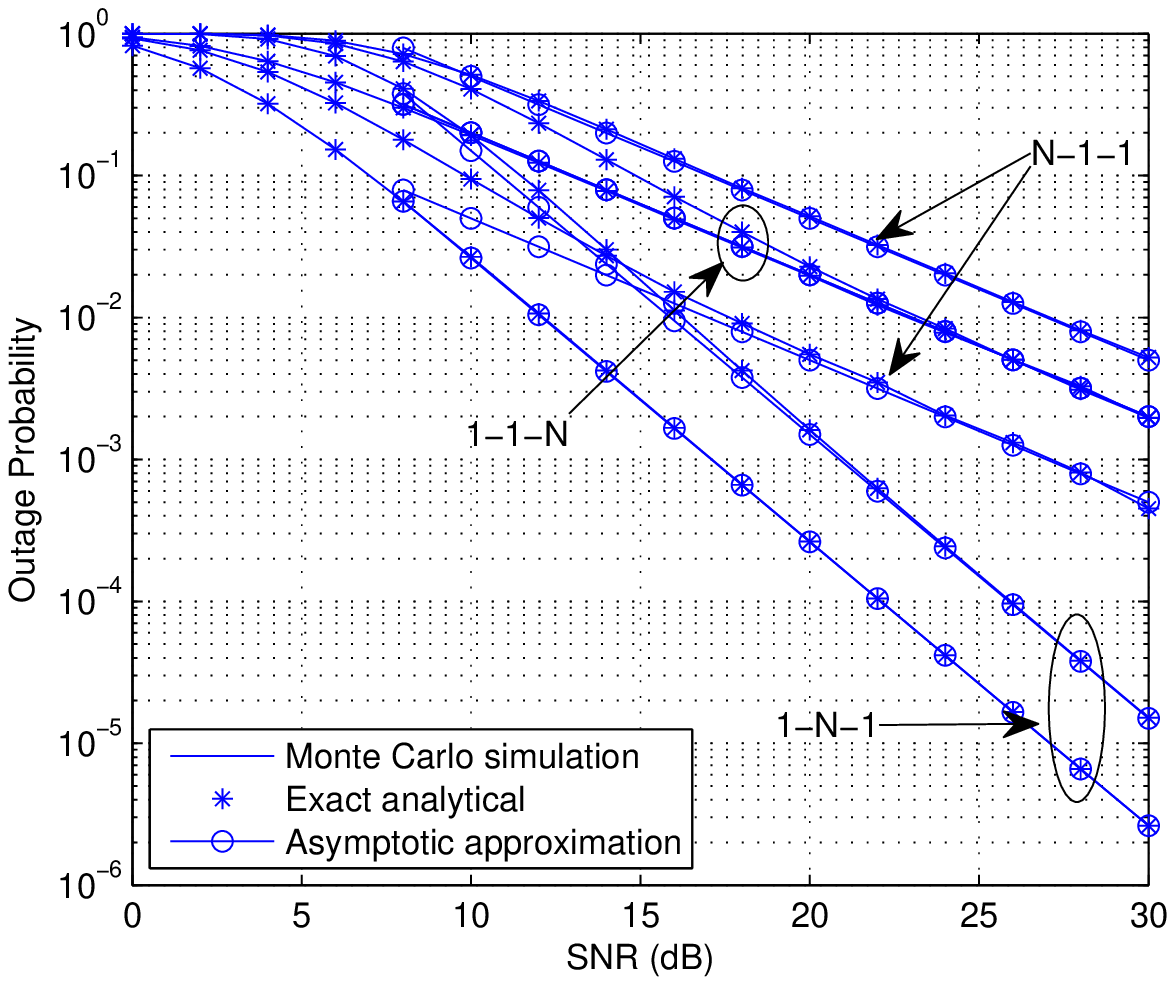}
\caption{Comparison of the outage probability of N-1-1, 1-1-N and 1-N-1 dual-hop systems with variable-gain
relaying.}\label{fig:fig5}
\end{figure}

\end{document}